\documentclass[12pt, draftclsnofoot, onecolumn]{IEEEtran}
\IEEEoverridecommandlockouts
\usepackage{cite}
\usepackage{amsmath,amssymb,amsfonts}
\usepackage{graphicx}
\usepackage{textcomp}
\usepackage{xcolor}
\usepackage{balance}
\usepackage{cite}
\usepackage{hyperref}
\usepackage{amsmath,amssymb,amsfonts}
\usepackage{amsmath}
\usepackage{amsthm}

\DeclareMathOperator*{\argmin}{argmin}
\usepackage{graphicx}
\usepackage{textcomp}
\usepackage{xcolor}
\usepackage{graphicx}
\usepackage{float}
\usepackage{subfigure}
\usepackage{amsmath}
\usepackage{amsfonts,amssymb,yfonts}
\usepackage{mathrsfs}
\usepackage{mathtools}
\usepackage{bm}
\usepackage{multirow}
\usepackage{array}
\usepackage{amssymb}
\usepackage{amsmath}
\usepackage{cite}
\usepackage{url}
\usepackage{xcolor}
\usepackage{cite,graphicx,amsmath,amssymb}
\usepackage{subfigure}
\usepackage{fancyhdr}
\usepackage{mdwmath}
\usepackage{mdwtab}
\usepackage{caption}
\usepackage{amsthm}
\usepackage{setspace}
\usepackage{bm}
\usepackage{mathtools}
\usepackage{dsfont}
\usepackage{bbm}
\usepackage{algorithm}
\usepackage{algorithmic}

\newtheorem{theorem}{Theorem}

\newtheorem{lemma}{Lemma}

\newtheorem{corollary}{Corollary}

\DeclareMathAlphabet\mathbffrak{OMS}{cmsy}{b}{n}
\makeatletter
\newcommand{\biggg}{\bBigg@{3}}
\newcommand{\Biggg}{\bBigg@{3.5}}
\makeatother
\def\BibTeX{{\rm B\kern-.05em{\sc i\kern-.025em b}\kern-.08em
    T\kern-.1667em\lower.7ex\hbox{E}\kern-.125emX}}
\begin{document}

\title{Sum-Rate Maximization for Movable Antenna Enabled Multiuser Communications}

\author{Zhenqiao~Cheng, Nanxi~Li, Jianchi~Zhu, and Chongjun~Ouyang
\thanks{Z. Cheng, N. Li, and J. Zhu are with the 6G Research Centre, China Telecom Beijing Research Institute, Beijing, 102209, China (e-mail: \{chengzq, linanxi, zhujc\}@chinatelecom.cn).}
\thanks{C. Ouyang is with the School of Information and Communication Engineering, Beijing University of Posts and Telecommunications, Beijing, 100876, China (e-mail: DragonAim@bupt.edu.cn).}
}
\maketitle

\begin{abstract}
A novel multiuser communication system with movable antennas (MAs) is proposed, where the antenna position optimization is exploited to enhance the downlink sum-rate. The joint optimization of the transmit beamforming vector and transmit MA positions is studied for a multiuser multiple-input single-input system. An efficient algorithm is proposed to tackle the formulated non-convex problem via capitalizing on fractional programming, alternating optimization, and gradient descent methods. To strike a better performance-complexity trade-off, a zero-forcing beamforming-based design is also proposed as an alternative. Numerical investigations are presented to verify the efficiency of the proposed algorithms and their superior performance compared with the benchmark relying on conventional fixed-position antennas (FPAs).
\end{abstract}

\begin{IEEEkeywords}
Antenna position, movable antenna (MA), multiuser communications, sum-rate maximization.
\end{IEEEkeywords}

\section{Introduction}
Multiple-input multiple-output (MIMO) technology stands as a cornerstone in the realm of wireless communications. It leverages multiple transceiving antennas to introduce an increased number of degrees of freedom (DoFs) into the wireless channel, thereby augmenting its spectral efficiency (SE) \cite{Heath2018}. However, traditional MIMO systems feature antennas that are immobile, rendering them unable to fully exploit the spatial dynamics inherent to wireless channels within a specific transmit/receive area. This limitation becomes particularly pronounced when the number of antennas is constrained \cite{Zhu2023}.

To harness additional spatial DoFs and further enhance SE, the concept of movable antennas (MAs) was conceived \cite{Zhu2022}. MAs are designed to overcome the constraints of conventional fixed-position antennas (FPAs). They achieve this by interfacing with radio frequency (RF) chains through flexible cables and incorporating real-time adjustability via controllers such as stepper motors or servos \cite{Ismail1991,Basbug2017}. This newfound flexibility empowers MAs to dynamically adapt their positions, effectively reshaping the wireless channel to deliver vastly improved wireless transmission capabilities \cite{Zhu2023}.

Owning to its superiority, the concept of MAs has garnered increasing research attention. The capacity of a point-to-point MIMO channel with MAs was initially characterized in \cite{Ma2022}. Subsequently, this research was extended to the uplink multiuser channel, wherein each user terminal (UT) is equipped with an MA, as explored in \cite{Zhu2023_2,Xiao2023,Pi2023}. In contrast to these prior contributions, our letter proposes a novel approach that harnesses the joint optimization of transmit beamforming and MA positions to enhance the sum-rate of a downlink multiuser multiple-input single-output (MU-MISO) system, with MAs
equipped at the base station (BS). It is worth noting that the authors of \cite{Wu2023} also investigated an MA-enabled MU-MISO system. However, their primary objective was to minimize the transmit power while guaranteeing the minimal rate requirement of each UT. Moreover, the MA elements in \cite{Wu2023} were constrained to move within a predetermined discrete grid, effectively rendering the MA-based system in \cite{Wu2023} equivalent to an FPA-based system with antenna selection. These distinctions underscore the uniqueness of the problem addressed in our letter compared to the one tackled in \cite{Wu2023}.

Our primary contributions are summarized as follows: {\romannumeral1}) We propose an MA-enabled downlink MU-MISO transmission framework that harnesses the MAs to optimize antenna positions for sum-rate improvements. {\romannumeral3}) We propose an efficient fractional programming (FP)-based algorithm to tackle the joint optimization of transmit beamforming and MA positions. {\romannumeral3}) We also propose a zero-forcing (ZF)-based design method to alleviate the complexity. {\romannumeral4}) Numerical results demonstrate that the proposed MA-based transmission provides more DoFs for improving the sum-rate than conventional FPA-based ones.
\begin{figure}[!t]
\centering
    \subfigbottomskip=-5pt
	\subfigcapskip=-5pt
\setlength{\abovecaptionskip}{0pt}
\includegraphics[height=0.35\textwidth]{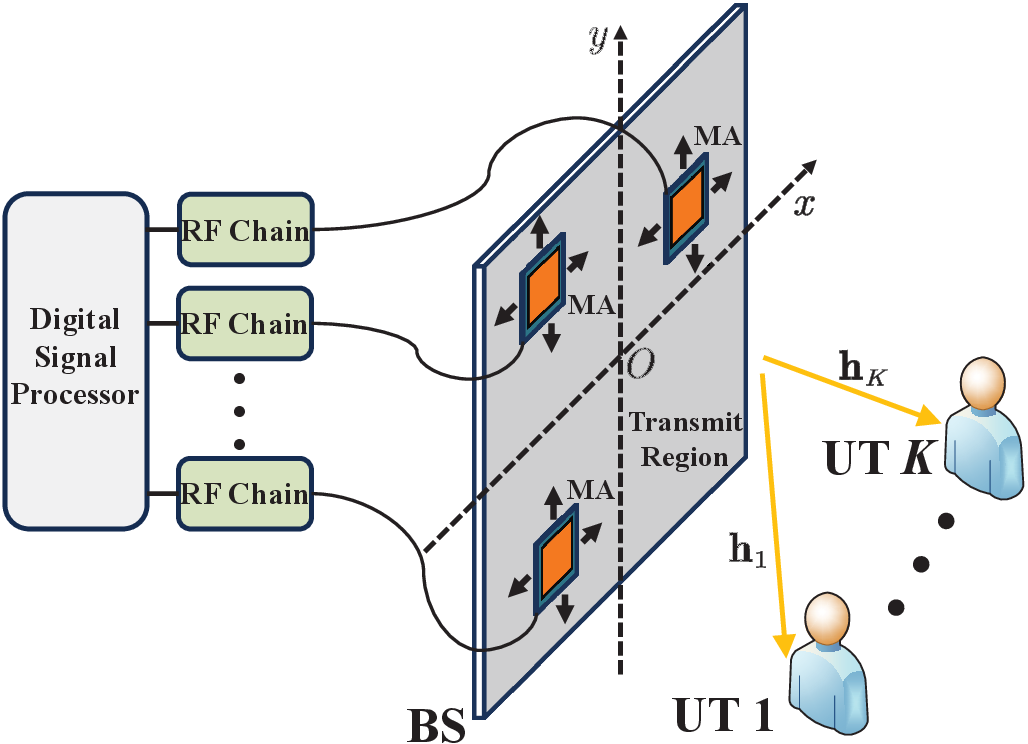}
\caption{The MA-enabled multiuser communication system}
\label{System_Model_MISOSE}
\vspace{-20pt}
\end{figure}
\section{System Model}
\subsection{System Description}
We consider MU-MISO transmission in an MA-enabled setting as depicted in {\figurename} {\ref{System_Model_MISOSE}}, where the BS simultaneously transmits signals to a set of $K$ single-antenna
UTs, which is denoted by ${\mathcal{K}}=\{1,\ldots,K\}$. The BS has $N$ transmit MAs and each UT $k\in{\mathcal{K}}$ has a single receive FPA. The MAs are connected to RF chains via flexible cables, and thus their positions can be adjusted in real time \cite{Ismail1991,Basbug2017}. The positions of the $n$th MA can be represented by Cartesian coordinates ${\mathbf{t}}_n=[x_n, y_n]^{\mathsf{T}}\in{\mathcal{C}}$ for $n\in{\mathcal{N}}=\{1,\ldots,N\}$, where $\mathcal{C}$ denotes the given two-dimensional region within which the MAs can move freely. Without loss of generality, we set $\mathcal{C}$ as square regions with size $A\times A$ \cite{Ma2022}.

We assume quasi-static block-fading channels, and focus on one particular fading block with the multi-path channel components at any location in $\mathcal{C}$ given as fixed. Denote the collections of the coordinates of $N$ MAs by ${\mathbf{T}}=[{\mathbf{t}}_1 \ldots {\mathbf{t}}_N]\in{\mathbbmss{R}}^{2\times N}$. The MISO propagations are described by the field-response based channel model \cite{Ma2023}, where the channel vector ${\mathbf{h}}_{k}\in{\mathbbmss{C}}^{N\times1}$ from the BS to UT $k$ follows the structure as 
{\setlength\abovedisplayskip{2pt}
\setlength\belowdisplayskip{2pt}
\begin{align}\label{Channel_Model}
{\mathbf{h}}_{k}={\mathbf{G}}_k^{\mathsf{H}}{\bm\Sigma}_k{\mathbf{1}}.
\end{align}
}The terms appearing in \eqref{Channel_Model} are defined as follows:
\begin{itemize}
  \item ${\mathbf{1}}\in{\{1\}}^{L_k\times1}$ is the all-one field response vector (FRV) at UT $k$, where $L_k$ is the number of channel paths.
  \item ${\bm\Sigma}_k={\mathsf{diag}}\{[\sigma_{k,1},\ldots,\sigma_{k,L_k}]^{\mathsf{T}}\}\in{\mathbbmss{C}}^{L_k\times L_k}$, where $\sigma_{k,\ell}$ is the complex response of the $\ell$th path for $\ell=1,\ldots,L_k$.
  \item ${\mathbf{G}}_k=[{\mathbf{g}}_{k,1}\ldots{\mathbf{g}}_{k,N}]\in{\mathbbmss{C}}^{L_k\times N}$ is the transmit FRV at the BS, where ${\mathbf{g}}_{k,n}\in{\mathbbmss{C}}^{L_k\times 1}$ is the transmit FRV between UT $k$ and the $n$th MA for $n=1,\ldots,N$.
  \item ${\mathbf{g}}_{k,n}=[{\rm{e}}^{{\rm{j}}\frac{2\pi}{\lambda}{\mathbf{t}}_n^{\mathsf{T}}{\bm\rho}_{k,1}},\ldots,{\rm{e}}^{{\rm{j}}\frac{2\pi}{\lambda}{\mathbf{t}}_n^{\mathsf{T}}{\bm\rho}_{k,L_k}}]^{\mathsf{T}}$, where ${\bm\rho}_{k,\ell}=[\sin{\theta_{k,\ell}}\cos{\phi_{k,\ell}},\cos{\theta_{k,\ell}}]^{\mathsf{T}}$, $\theta_{k,\ell}\in[0,\pi]$ and $\phi_{k,\ell}\in[0,\pi]$ are the elevation and azimuth angles of the $\ell$th path, respectively, and $\lambda$ is the wavelength.
\end{itemize}
Taken together, we have ${\mathbf h}_{k}=[h_{k}({\mathbf{t}}_1)\ldots h_{k}({\mathbf{t}}_N)]^{\mathsf{T}}$, where 
{\setlength\abovedisplayskip{2pt}
\setlength\belowdisplayskip{2pt}
\begin{align}\label{Reponse_Model}
h_{k}({\mathbf{t}})\triangleq\sum\nolimits_{\ell=1}^{L_k}\sigma_{k,\ell}{\rm{e}}^{-{\rm{j}}\frac{2\pi}{\lambda}{\mathbf{t}}^{\mathsf{T}}{\bm\rho}_{k,\ell}}.
\end{align}
}It is worth noting that the MA-based channel vectors are determined by the signal propagation environment and the positions of MAs. The system operates in the time division duplexing (TDD) mode. The channel state information (CSI) is hence estimated in the uplink training phase via pilot sequences. We assume that the pilots are mutually orthogonal and that the estimation error is negligible. The BS thus learns perfectly the CSI. Details on channel estimation for MA-aided communication systems are found in \cite{Ma2023}.

Denoted by ${\mathbf{x}}\in{\mathbbmss{C}}^{N\times1}$ the input of the downlink transmission, the received signal of UT $k$ is expressed as follows:
{\setlength\abovedisplayskip{2pt}
\setlength\belowdisplayskip{2pt}
\begin{align}\label{Signal_Model}
y_k=\mathbf{h}_k^{\mathsf{H}}{\mathbf{x}}+n_k,
\end{align}
}where $n_k$ denotes the circularly symmetric complex Gaussian noise with zero mean and covariance $\sigma_k^2$. Note that ${\mathbf{x}}=\sum_{k=1}^{K}{\mathbf{w}}_kx_k$, where $x_k\in{\mathbbmss{C}}$ is the data symbol intended for UT $k$ with ${\mathbf{w}}_k\in{\mathbbmss{C}}^{N\times1}$ being its transmit beamforming vector. In addition, the data symbol $x_k$ $(\forall k)$ is considered to have zero mean and unit
variance. 
Hence, the decoding signal-to-interference-plus-noise ratio (SINR) of $s_k$ at UT $k$ is given by $\gamma_k=\frac{\lvert\mathbf{h}_k^{\mathsf{H}}{\mathbf{w}}_{k}\rvert^2}{\sum_{k'\ne k}\lvert\mathbf{h}_k^{\mathsf{H}}{\mathbf{w}}_{k'}\rvert^2+\sigma_k^2}$. The sum-rate can be expressed as ${\mathcal{R}}=\sum\nolimits_{k=1}^{K}\log(1+\gamma_k)$. Note that different from the conventional multiuser channel with FPAs, the sum-rate for the MA-enabled multiuser channel, i.e., ${\mathcal{R}}$, depends on the positions of MAs $\mathbf{T}$, which influence the channel vectors $\{{\mathbf{h}}_k\}_{k\in{\mathcal{K}}}$ and the beamformer ${\mathbf{W}}=[{\mathbf{w}}_1\ldots{\mathbf{w}}_K]\in{\mathbbmss{C}}^{N\times K}$.
\subsection{Problem Formulation}
In order to avoid the coupling effect between the antennas in the transmit region, a minimum distance $D$ is required between each pair of antennas, i.e., $\lVert {\mathbf{t}}_n-{\mathbf{t}}_{n'}\rVert\leq D$ for $n\ne n'$ \cite{Ma2022}. Then, we aim to improve the sum-rate by jointly optimizing the MA positions $\mathbf{T}$ and the transmit beamformer $\mathbf{W}$. The optimization problem is formulated as follows:
{\setlength\abovedisplayskip{2pt}
\setlength\belowdisplayskip{2pt}
\begin{subequations}\label{Sum_Rate_Max_Problem}
\begin{align}
{{\mathcal{P}}_1}:&\max\nolimits_{\mathbf{T},{\mathbf{W}}}~{\mathcal{R}}\\
&{\rm{s.t.}}~{\mathsf{tr}}({\mathbf{WW}}^{\mathsf{H}})\leq p,\label{Sum_Rate_Max_C1}\\
&\quad~~{\mathbf{t}}_n\in{\mathcal{C}},\forall n\in{\mathcal{N}},\lVert {\mathbf{t}}_n-{\mathbf{t}}_{n'}\rVert\leq D,n\ne n',\label{Sum_Rate_Max_C2}
\end{align}
\end{subequations}
}where $p$ is the power budget. Note that ${{\mathcal{P}}_1}$ is a non-convex optimization problems due to the non-convexity of $\mathcal{R}$ with respect to (w.r.t.) $({\mathbf{T}},{\mathbf{W}})$ and the non-convex minimum distance constraint $\lVert {\mathbf{t}}_n-{\mathbf{t}}_{n'}\rVert\leq D$. Moreover, the beamformer $\mathbf{W}$ is coupled with $\mathbf{T}$, which makes ${{\mathcal{P}}_1}$ challenging to solve.
\section{Proposed Solution}
In this section, we present a pair of efficient algorithms to solve problem ${{\mathcal{P}}_1}$. First, ${{\mathcal{P}}_1}$ is simplified into a more tractable yet equivalent form w.r.t. $\{{\mathbf{W}}\}\cup\{\mathbf{t}_n\}_{n=1}^{N}$ by invoking the FP framework \cite{Shen2018}. Then, the beamforming matrix $\mathbf{W}$ and each MA position ${\mathbf{t}}_n$ are updated in an alternating manner, with all the other variables being fixed. After that, by deriving a more tractable expression for the sum-rate achieved by the ZF beamforming, we present an alternative solution for problem ${{\mathcal{P}}_1}$ with lower complexity.
\subsection{FP-Based Design}
\subsubsection{Reformulation of ${{\mathcal{P}}_1}$}
Invoking the FP technique, i.e., the Lagrangian dual transform and quadratic transform methods \cite{Shen2018}, we introduce two auxiliary variable ${\bm\lambda}=[\lambda_1,\ldots,\lambda_K]$ and ${\bm\beta}=[\beta_1,\ldots,\beta_K]$ to derive the lemma as follows.
\vspace{-5pt}
\begin{lemma}\label{Lemma_FP}
Problem ${{\mathcal{P}}_1}$ in \eqref{Sum_Rate_Max_Problem} is equivalent to
{\setlength\abovedisplayskip{2pt}
\setlength\belowdisplayskip{2pt}
\begin{subequations}\label{Sum_Rate_Maximization_Problem_FP}
\begin{align}
&{\mathcal{P}}_2:\max_{\mathbf{T},{\mathbf{W}},{\bm\lambda},{\bm\beta}}~{\mathcal{L}}=\sum\nolimits_{k=1}^{K}\log(1+\lambda_k)-\sum\nolimits_{k=1}^{K}\lambda_k\nonumber\\
&\qquad\qquad+\sum\nolimits_{k=1}^{K}(1+\lambda_k)[2\Re\{\beta_k^{*}a_{k}\}-\lvert\beta_k\rvert^2 B_k]\\
&{\rm{s.t.}}~\eqref{Sum_Rate_Max_C1},\eqref{Sum_Rate_Max_C2},\lambda_k>0,\beta_k\in{\mathbbmss{C}},k\in{\mathcal{K}},
\end{align}
\end{subequations}
}where the optimal values of $\lambda_k$ and $\beta_k$ are given by $\lambda_k^{\star}=\lvert a_k\rvert^2/(\sum_{k'\ne k}\lvert\mathbf{w}_{k'}^{\mathsf{H}}{\mathbf{h}}_{k}\rvert^2+\sigma_k^2)=\gamma_k$ and $\beta_k^{\star}=a_kB_k^{-1}$, respectively, with $a_k={\mathbf{w}}_k^{\mathsf{H}}{\mathbf{h}}_k$ and $B_k=\sigma_k^2+\sum_{i=1}^{K}\lvert{\mathbf{w}}_i^{\mathsf{H}}{\mathbf{h}}_k\rvert^2$.
\end{lemma}
\vspace{-5pt}
\begin{IEEEproof}
Please refer to \cite{Shen2018} for more details.
\end{IEEEproof}
To decouple the variables $\{\mathbf{T},{\mathbf{W}},{\bm\lambda},{\bm\beta}\}$ in \eqref{Sum_Rate_Maximization_Problem_FP}, we propose to optimize each variable iteratively with other variables fixed. Since the conditionally optimal $\bm\lambda$ and $\bm\beta$ are already presented in Lemma \ref{Lemma_FP}, we propose to develop the iterative design of $\{{\mathbf{W}},{\mathbf{t}}_1,\ldots,{\mathbf{t}}_N\}$ with given optimal $\{{\bm\lambda},{\bm\beta}\}$.
\subsubsection{Optimizing the Beamforming Matirx $\mathbf{W}$}
The marginal problem for $\mathbf{W}$ is expressed as follows:
{\setlength\abovedisplayskip{2pt}
\setlength\belowdisplayskip{2pt}
\begin{align}
{\mathbf W}^{\star}\!=\!\argmin\nolimits_{{\mathsf{tr}}({\mathbf{W}}{\mathbf{W}}^{\mathsf H})\leq p}({\mathsf{tr}}({\mathbf W}^{\mathsf H}{\mathbf C}{\mathbf W})\!-\!2\Re\{{\mathsf{tr}}({\mathbf W}^{\mathsf H}{\mathbf D})\}),\nonumber
\end{align}
}where ${\mathbf C}=\sum_{k=1}^{K}(1+\lambda_k)\lvert\beta_k\rvert^2{\mathbf{h}}_k{\mathbf{h}}_k^{\mathsf{H}}\in{\mathbbmss{C}}^{N\times N}$ and ${\mathbf D}=[(1+\lambda_1)\beta_1^{*}{\mathbf{h}}_1\ldots(1+\lambda_K)\beta_K^{*}{\mathbf{h}}_K]\in{\mathbbmss{C}}^{N\times K}$. This is a standard convex quadratic optimization problem whose solution is \cite{Boyd2004}
{\setlength\abovedisplayskip{2pt}
\setlength\belowdisplayskip{2pt}
\begin{align}\label{Optimal_Precoding}
{\mathbf W}^{\star}=({\mathbf{C}}+\lambda{\mathbf I})^{-1}{\mathbf{D}}.
\end{align}
}The regularizer $\lambda$ is chosen, such that the complementarity slackness condition, i.e., $\lambda({\mathsf{tr}}({\mathbf{W}}{\mathbf{W}}^{\mathsf H})- p)=0$, is satisfied. If ${\mathsf{tr}}({\mathbf{D}}^{\mathsf H}({\mathbf{C}}+\lambda{\mathbf I})^{-2}{\mathbf{D}})=p$; then, $\lambda=0$. Otherwise, we can obtain the solution of $\lambda$ from the following identity:
{\setlength\abovedisplayskip{2pt}
\setlength\belowdisplayskip{2pt}
\begin{align}
{\mathsf{tr}}({\mathbf{W}}{\mathbf{W}}^{\mathsf H})={\mathsf{tr}}({\mathbf{D}}^{\mathsf H}({\mathbf{C}}+\lambda{\mathbf I})^{-2}{\mathbf{D}})=p.
\end{align}
}Denote the eigen-decomposition of ${\mathbf C}$ as ${\mathbf U}^{\mathsf H}{\bm\Lambda}{\mathbf U}$ yields
{\setlength\abovedisplayskip{2pt}
\setlength\belowdisplayskip{2pt}
\begin{align}
\sum\nolimits_{n=1}^{N}\frac{\lvert[{\mathbf U}{\mathbf D}{\mathbf D}^{\mathsf H}{\mathbf U}^{\mathsf H}]_{n,n}\rvert^2}{([{\bm\Lambda}]_{n,n}+\lambda)^{2}}=p,\label{Power_Bisection}
\end{align}
}where $[\mathbf{V}]_{i,j}$ is the $(i,j)$th element of matrix $\mathbf{V}$. Since $[{\bm\Lambda}]_{n,n}\geq0$ for $n\in{\mathcal{N}}$, the left-hand side of \eqref{Power_Bisection} is a monotonic function with $\lambda\geq0$. Thus, we can find $\lambda$ by solving equation \eqref{Power_Bisection} via a bisection-based search.
\subsubsection{Optimizing the MA Position ${\mathbf{t}}_n$}
The marginal problem for ${\mathbf{t}}_n$ is expressed as follows:
{\setlength\abovedisplayskip{2pt}
\setlength\belowdisplayskip{2pt}
\begin{align}\label{MA_Subproblem}
\max_{{\mathbf{t}}_n\in{\mathcal{S}}_n}f_n({\mathbf{t}}_n)\!\triangleq\!\sum_{k=1}^{K}(2\Re\{h_k^{*}({\mathbf{t}}_n)c_{k,n}\}\!-\!d_{k,n}\lvert h_k({\mathbf{t}}_n)\rvert^2),
\end{align}
}where $d_{k,n}=(1+\lambda_k)\lvert\beta_k\rvert^2W_{n,n}$, $c_{k,n}=(1+\lambda_k)(\beta_kw_{k,n}-\lvert\beta_k\rvert^2\sum_{n'\ne n}W_{n,n'}h_k({\mathbf{t}}_{n'}))$, $w_{k,n}$ is the $n$th elements of ${\mathbf{w}}_k$, $W_{i,j}$ is the $(i,j)$th element of matrix $\sum_{k=1}^{K}{\mathbf{w}}_k{\mathbf{w}}_k^{\mathsf{H}}$, and ${\mathcal{S}}_n\triangleq\{{\mathbf{x}}|{\mathbf{x}}\in{\mathcal{C}},\lVert {\mathbf{x}}-{\mathbf{t}}_{n'}\rVert\leq D,\forall n\ne n'\}$. Due to the intractability of $f_n(\cdot)$, stationary points of subproblem \eqref{MA_Subproblem} can be found capitalizing on the gradient decent method with backtracking line search \cite{Boyd2004}. To this end, the gradient values of $f_n({\mathbf{t}}_n)$ w.r.t. ${\mathbf{t}}_n$ are calculated as follows: \cite{Gesbert2007}
{\setlength\abovedisplayskip{2pt}
\setlength\belowdisplayskip{0pt}
\begin{equation}\label{Der_MA}
\begin{split}
&\nabla_{{\mathbf{t}}_n}f_{n}=\sum_{k=1}^{K}\!\sum_{\ell=1}^{L_i}\!\frac{\lvert\tau_{n}^{k,\ell}\rvert}{-\frac{\lambda}{4\pi}}\sin\!\left(\!\frac{2\pi}{\lambda}
{\mathbf{t}}_n^{\mathsf{T}}{\bm\rho}_{k,\ell}\!+\!\angle\tau_{n}^{k,\ell}\!\right)\!{\bm\rho}_{k,\ell}\\
&+\!\sum_{k=1}^{K}\!\sum_{\ell=1}^{L_k}\!\sum_{\ell'\ne \ell}\!\frac{\lvert\sigma_{\ell,k}\sigma_{\ell',k}\rvert}{\frac{\lambda}{4\pi d_{k,n}}}\sin\!\left(\!\frac{2\pi}{\lambda}
{\mathbf{t}}_n^{\mathsf{T}}{\bm\rho}_k^{\ell,\ell'}\!+\!\theta_k^{\ell,\ell'}\!\right)\!{\bm\rho}_k^{\ell,\ell'},
\end{split}
\end{equation}
}where $\tau_{n}^{k,\ell}=\sigma_{k,\ell}^{*}c_{k,n}$, ${\bm\rho}_k^{\ell,\ell'}={\bm\rho}_{k,\ell}-{\bm\rho}_{k,\ell'}$, and $\theta_k^{\ell,\ell'}=\angle{\sigma}_{\ell',k}-\angle{\sigma}_{\ell,k}$. The algorithm for optimizing $\mathbf{T}$ is given in Algorithm \ref{Algorithm1}. Since the sum-rate is upper bounded, the convergence is guaranteed. Regarding the complexity of Algorithm \ref{Algorithm1}, it scales with ${\mathcal{O}}(IN(\sum_{k=1}^{K}L_k^2)\log_2\frac{1}{u_{\min}})$, where $I$ is the number of iterations and $u_{\min}$ denotes the accuracy.
\begin{algorithm}[!t]
  \algsetup{linenosize=\tiny} \scriptsize
  \caption{Gradient-Based Algorithm for Optimizing $\mathbf{T}$}
  \label{Algorithm1}
  \begin{algorithmic}[1]
    \STATE Initialize ${\mathbf{T}}^{0}=[{\mathbf{t}}_1^0 \ldots {\mathbf{t}}_N^0]$, the maximum iteration number $I$, step size $u_{\rm{ini}}$, the minimum tolerance step size $u_{\min}$, and set the current iteration $a=0$;
    \REPEAT
    \FORALL{$n=1:N$} 
    \STATE Compute the gradient value $\nabla_{{\mathbf{t}}_n^{a}}{f_n}$ and set $u=u_{\rm{ini}}$;
      \REPEAT
      \STATE Compute $\hat{\mathbf{t}}_n={\mathbf{t}}_n^{a}+u\cdot\nabla_{{\mathbf{t}}_n^{a}}{f_n}$ and set $u=u/2$;     
      \UNTIL{$\hat{\mathbf{t}}_n\in{\mathcal{S}}_n \& f_n(\hat{\mathbf{t}}_n)>f_n({\mathbf{t}}_n^{a})$ or $u<u_{\min}$};
      \STATE Set ${\mathbf{t}}_n^{a}=\hat{\mathbf{t}}_n$ and update ${\mathbf{t}}_n^{a+1}=\hat{\mathbf{t}}_n$;
    \ENDFOR
      \STATE Update $a=a+1$;
    \UNTIL{convergence or the maximum iteration number $I$ is reached}.
  \end{algorithmic}
\end{algorithm} 
\begin{algorithm}[!t]
  \algsetup{linenosize=\tiny} \scriptsize
  \caption{FP-Based Algorithm for Solving Problem ${{\mathcal{P}}_1}$}
  \label{Algorithm2}
  \begin{algorithmic}[1]
    \STATE Initialize $\{{\mathbf{W}}^{0},{\mathbf{T}}^{0}\}$, the maximum iteration number $I_{\rm{fp}}$, and set the current iteration $t=0$;
    \REPEAT
      \STATE Update ${\bm\lambda}^{t}$ and ${\bm\beta}^{t}$ based on Lemma \ref{Lemma_FP};
      \STATE Optimize $\mathbf{W}$ for given $\{{\bm\lambda}^{t},{\bm\beta}^{t},{\mathbf{T}}^{t}\}$ by \eqref{Optimal_Precoding}, and obtain ${\mathbf{W}}^{t+1}$;
      \STATE Solve problem \eqref{MA_Subproblem} for given $\{{\bm\lambda}^{t},{\bm\beta}^{t},{\mathbf{W}}^{t+1}\}$ by applying the gradient decent method summarized in Algorithm \ref{Algorithm1}, and obtain ${\mathbf{T}}^{t+1}$;
      \STATE Update $t=t+1$;
    \UNTIL{convergence or the maximum iteration number $I_{\rm{fp}}$ is reached}.
  \end{algorithmic}
\end{algorithm} 
\subsubsection{Convergence and Complexity Analyses}
The derived FP-based algorithm is summarized in Algorithm \ref{Algorithm2}, which is guaranteed to converge to a stationary solution of problem ${{\mathcal{P}}_1}$ \cite{Shen2018}. The computational complexity of the proposed algorithm can be further characterized in terms of problem dimensions. To this end, let $I_{\rm{fp}}$ denote the numbers of iterations. The per-iteration computational complexity is composed of the complexity of updating variables $\{{\bm\lambda},{\bm\beta},{\mathbf{W}},{\mathbf{T}}\}$. It is readily shown that the complexity of marginal optimizations w.r.t. ${\bm\lambda}$, $\bm\beta$, $\mathbf{W}$, and $\mathbf{T}$ scales with $\mathcal{O}(KN)$, $\mathcal{O}(KN)$, ${\mathcal{O}}(N^3)$, and ${\mathcal{O}}(IN(\sum_{k=1}^{K}L_k^2)\log_2\frac{1}{u_{\min}})$, respectively. Hence, the overall complexity of Algorithm \ref{Algorithm2} scales with $\mathcal{O}(I_{\rm{fp}}(2KN+N^3+IN(\sum_{k=1}^{K}L_k^2)\log_2\frac{1}{u_{\min}}))$, which is of a polynomial order.
\subsection{ZF-Based Design}\label{ZF_MA}
The preceding subsection introduced the FP-based algorithm, which alternates between optimizing the MA position matrix $\mathbf{T}$ and the beamforming matrix $\mathbf{W}$. While this approach maintains reasonable computational complexity, it may impose significant computational burdens in various practical applications. To address this concern, we present an alternative scheme rooted in ZF beamforming in this part. Here, our approach involves the initial design of a ZF-based beamformer ${\mathbf{W}}=\sqrt{\frac{p}{{\mathsf{tr}}(({\mathbf{H}}^{\mathsf{H}}{\mathbf{H}})^{-1})}}{\mathbf{H}}({\mathbf{H}}^{\mathsf{H}}{\mathbf{H}})^{-1}\triangleq{\mathbf{W}}_{\rm{ZF}}$, constructed from the channel matrix ${\mathbf{H}}=[{\mathbf{h}}_1\ldots{\mathbf{h}}_K]\in{\mathbbmss{C}}^{N\times K}$. Subsequently, we focus on optimizing the MA positions using the gradient descent method. The sum-rate achieved by ${\mathbf{W}}_{\rm{ZF}}$ is given by ${\mathcal{R}}=\sum_{k=1}^{K}\log(1+p/\sigma_k^2/{\mathsf{tr}}(({\mathbf{H}}^{\mathsf{H}}{\mathbf{H}})^{-1}))$. Thus, the subproblem of MA position optimization is formulated as follows:
{\setlength\abovedisplayskip{2pt}
\setlength\belowdisplayskip{2pt}
\begin{align}\label{P_ZF}
\max\nolimits_{\mathbf{T}}\tilde{f}(\mathbf{T})\triangleq-{\mathsf{tr}}(({\mathbf{H}}^{\mathsf{H}}{\mathbf{H}})^{-1})\quad{\rm{s.t.}}~\eqref{Sum_Rate_Max_C2}.
\end{align}
}

Despite the intractability of $\tilde{f}(\mathbf{T})$ and the tight coupling of $\{{\mathbf{t}}_n\}_{n=1}^{N}$, we could propose a suboptimal algorithm to tackle problem \eqref{P_ZF} with guaranteed convergence, capitalizing on the approaches of alternating optimization, gradient decent, and backtracking line search. To elucidate our approach, we initiate by partitioning the variable set $\mathbf{T}$ into $N$ distinct blocks $\{{\mathbf{t}}_n\}_{n=1}^{N}$. We then proceed to address $N$ subproblems of \eqref{P_ZF}, each of which optimizes a specific transmit MA position $\mathbf{t}_n$ while keeping all other variables fixed. The resulting alternating optimization algorithm efficiently iterates through these $N$ subproblems, gradually refining the solution to \eqref{P_ZF}. Employing complex-valued matrix differentiation principles \cite{Gesbert2007}, we compute the derivative of $\tilde{f}(\mathbf{T})$ w.r.t. $\mathbf{t}_n$ as follows:
{\setlength\abovedisplayskip{2pt}
\setlength\belowdisplayskip{0pt}
\begin{align}
\nabla_{{\mathbf{t}}_n}\tilde{f}&=\sum\nolimits_{k_1=1}^{K}\!\sum\nolimits_{k_2=1}^{K}\!\Big(H_{k_1,k_2}
\sum\nolimits_{\ell=1}^{L_{k_1}}\frac{{\rm{j}}2\pi\sigma_{k_1,\ell}^{*}h_{k_2}({\mathbf{t}}_n)}{\lambda{\rm{e}}^{-{\rm{j}}\frac{2\pi}{\lambda}{\mathbf{t}}_n^{\mathsf{T}}{\bm\rho}_{k_1,\ell}}}\nonumber\\
&\times{\bm\rho}_{k_1,\ell}+\sum\nolimits_{\ell=1}^{L_{k_2}}\frac{-{\rm{j}}2\pi\sigma_{k_2,\ell}h_{k_1}^{*}({\mathbf{t}}_n)}
{\lambda{\rm{e}}^{{\rm{j}}\frac{2\pi}{\lambda}{\mathbf{t}}_n^{\mathsf{T}}{\bm\rho}_{k_2,\ell}}}{\bm\rho}_{k_2,\ell}\Big),\nonumber
\end{align}
}where $H_{k_1,k_2}$ is the $(k_1,k_2)$th element of matrix $({\mathbf{H}}^{\mathsf{H}}{\mathbf{H}})^{-2}$. 

It is worth noting that the overall algorithm for solving problem \eqref{P_ZF} involves similar steps as Algorithm \ref{Algorithm1}, and the associated computational complexity scales with ${\mathcal{O}}(IN(K^3+K^2N+K\sum_{k=1}^{K}L_k)\log_2\frac{1}{u_{\min}})$. In contrast to the FP-based approach, the ZF-based design eliminates the need for alternating updates between $\mathbf{W}$ and $\mathbf{T}$, thus offering a desirable reduction in computational complexity. A more comprehensive comparison of the computational complexities of our proposed algorithms will be provided in Section \ref{Simulations}.
\section{Numerical Results}\label{Simulations}
In this section, numerical results are provided to validate the effectiveness of our proposed algorithms. In the simulation, we set $N = 4$, $K=4$, $D = \frac{\lambda}{2}$, $\sigma_k^2=-100$ dBm ($\forall k$), $u_{\min}=10^{-3}$, $u_{\rm{ini}}=10$, $I=20$, and $I_{\rm{fp}}=50$. The UTs are distributed uniformly over a hexagonal cell with a radius of 500 m. Moreover, we have incorporated the free-space path loss model for UT $k$, given by $-10\log_{10}{\mu_k}=92.5+20\log_{10}[f_0({\text{GHz}})]+20\log_{10}[d_k({\text{km}})]$, where $f_0 = 5$ GHz is the carrier frequency and $d_k$ is the distance between the BS and UT $k$. As for the channel model, we assume that $L_{k}=4$ ($\forall k$) and $\sigma_{l,k}\sim{\mathcal{CN}}(0,\frac{\mu_k}{L_{k}})$ ($\forall l,k$). The elevation and azimuth angles are randomly set within $[0,\pi]$. We compare the performance of our proposed algorithms with the FPA-based benchmark scheme, where the BS is equipped with an FPA-based uniform linear array with $N$ antennas spaced by $\frac{\lambda}{2}$. The presented numerical results are averaged over 1000 independent channel realizations with randomly initialized optimization variables. All simulations are conducted using Mathworks MATLAB R2020b on the computer with a 2.60-GHz i5-13500H CPU and 32-GB RAM.
\begin{figure}[!t]
    \centering
    \subfigbottomskip=0pt
	\subfigcapskip=-5pt
\setlength{\abovecaptionskip}{0pt}
    \subfigure[Number of iterations. $A=2\lambda$.]
    {
        \includegraphics[height=0.35\textwidth]{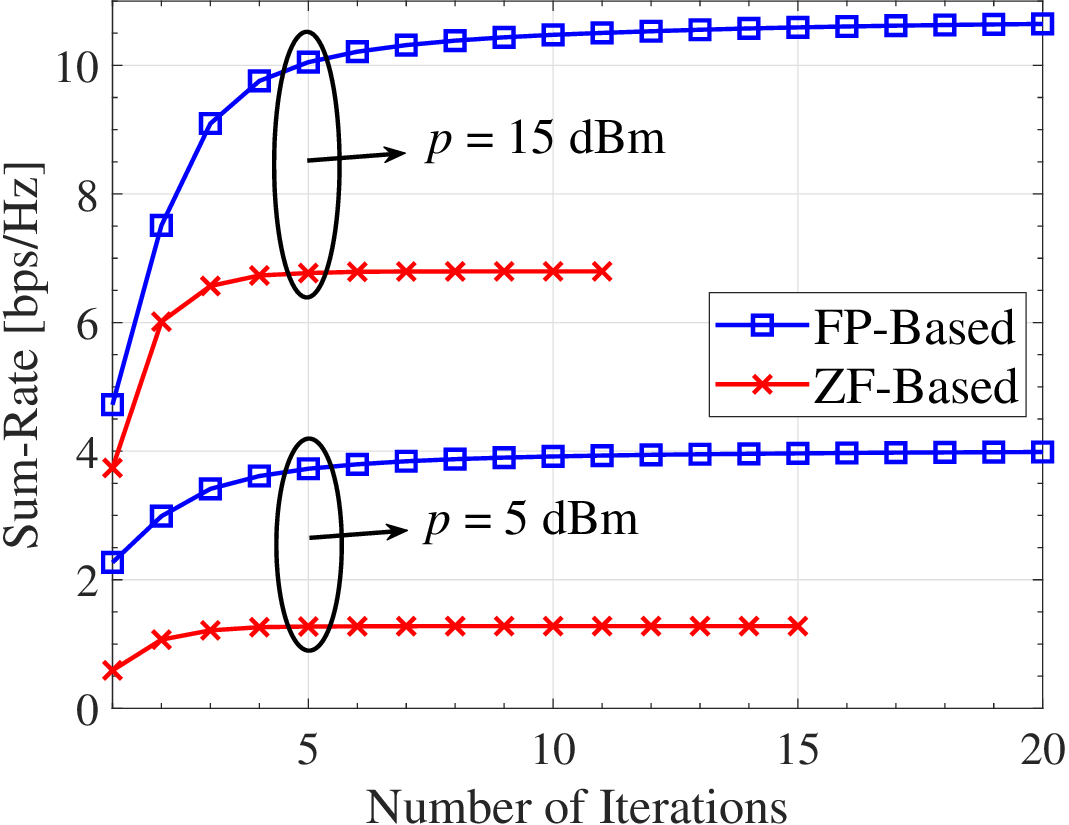}
	   \label{fig1a}	
    }
   \subfigure[Running time.]
    {
        \includegraphics[height=0.35\textwidth]{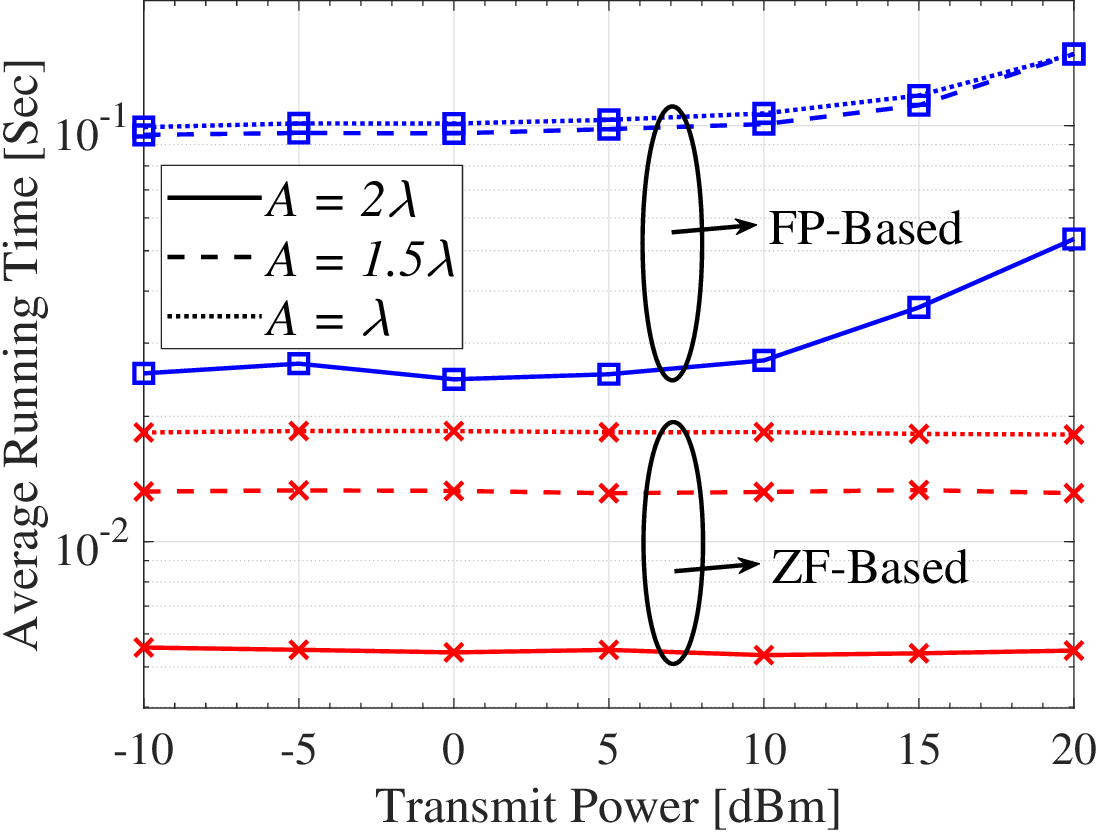}
	   \label{fig1b}	
    }
\caption{Complexity of the proposed algorithms for MAs.}
    \label{figure1}
    \vspace{-10pt}
\end{figure}

In {\figurename} {\ref{fig1a}}, we first depict the convergence behavior of our proposed algorithms for MAs. From {\figurename} {\ref{fig1a}}, it is observed that the sum-rate of the proposed algorithms increases quickly with the number of iterations. The proposed FP-based and ZF-based designs converge with around 12 and 5 iterations, respectively. In {\figurename} {\ref{fig1b}}, we compare the computational complexities of our proposed algorithms in terms of the CPU running time. It is observed that the ZF-based design requires less CPU time to achieve convergence when compared to the FP-based approach. The numerical results presented in {\figurename} {\ref{figure1}} suggest that the ZF-based design is more computationally efficient than the FP-based one, which is consistent with our previous arguments in Section \ref{ZF_MA}.

\begin{figure}[!t]
\centering
    \subfigbottomskip=0pt
	\subfigcapskip=-5pt
\setlength{\abovecaptionskip}{0pt}
\includegraphics[height=0.35\textwidth]{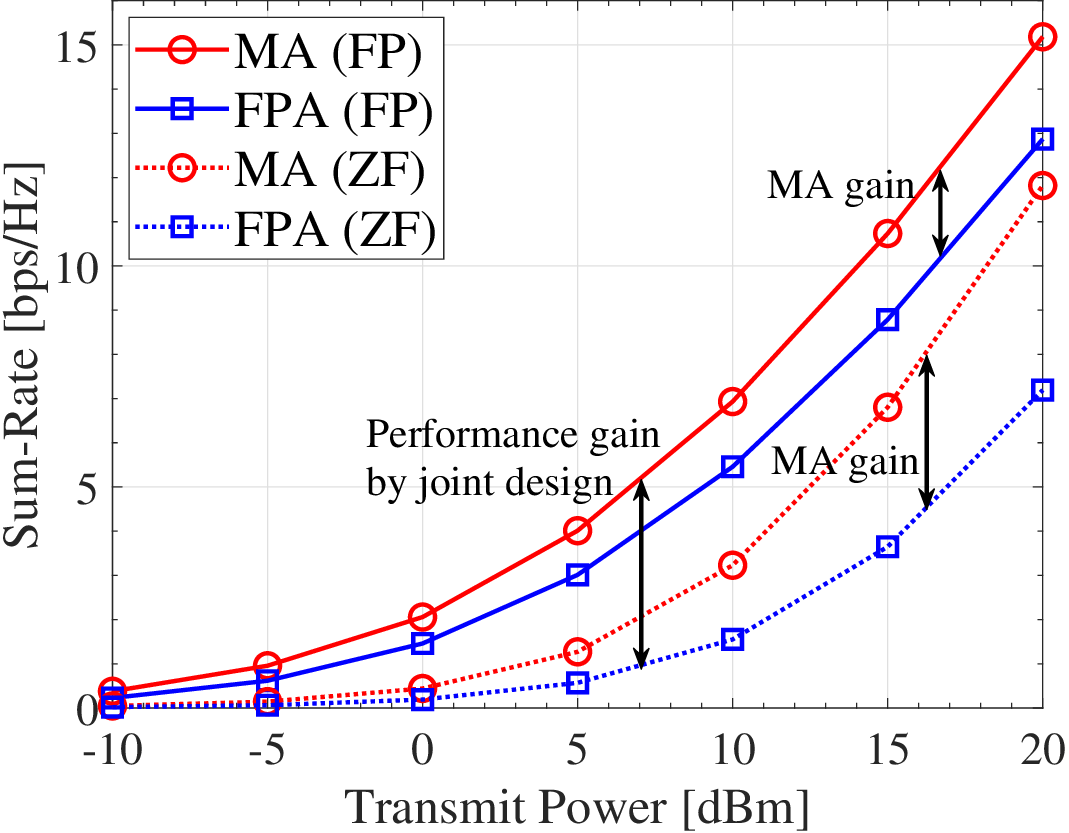}
\caption{Sum-rate vs. the power budget. $A=2\lambda$.}
\label{fig2a}
\vspace{-10pt}
\end{figure}

In {\figurename} {\ref{fig2a}}, we present the sum-rate of the proposed and benchmark schemes versus the transmit power $p$. It is observed that with the same power, our proposed algorithms can achieve a larger sum-rate as compared to the schemes with FPAs. For instance, when we consider the scenario with $p=5$ dBm, our proposed FP-based and ZF-based schemes exhibit notable performance improvements of 33.1\% and 123.3\%, respectively, over the FPA-based schemes. These substantial gains in sum-rate performance are primarily attributable to the optimization of MA positions. Furthermore, a noteworthy observation is that our proposed MA-based framework, when employing FP, achieves a sum-rate that surpasses that of the FPA-based scheme using ZF by a factor exceeding 6-fold. This observation underscores the superiority and efficacy of the joint optimization of transmit beamforming and MA positions in significantly enhancing the overall performance.

\begin{figure}[!t]
\centering
    \subfigbottomskip=0pt
	\subfigcapskip=-5pt
\setlength{\abovecaptionskip}{0pt}
\includegraphics[height=0.35\textwidth]{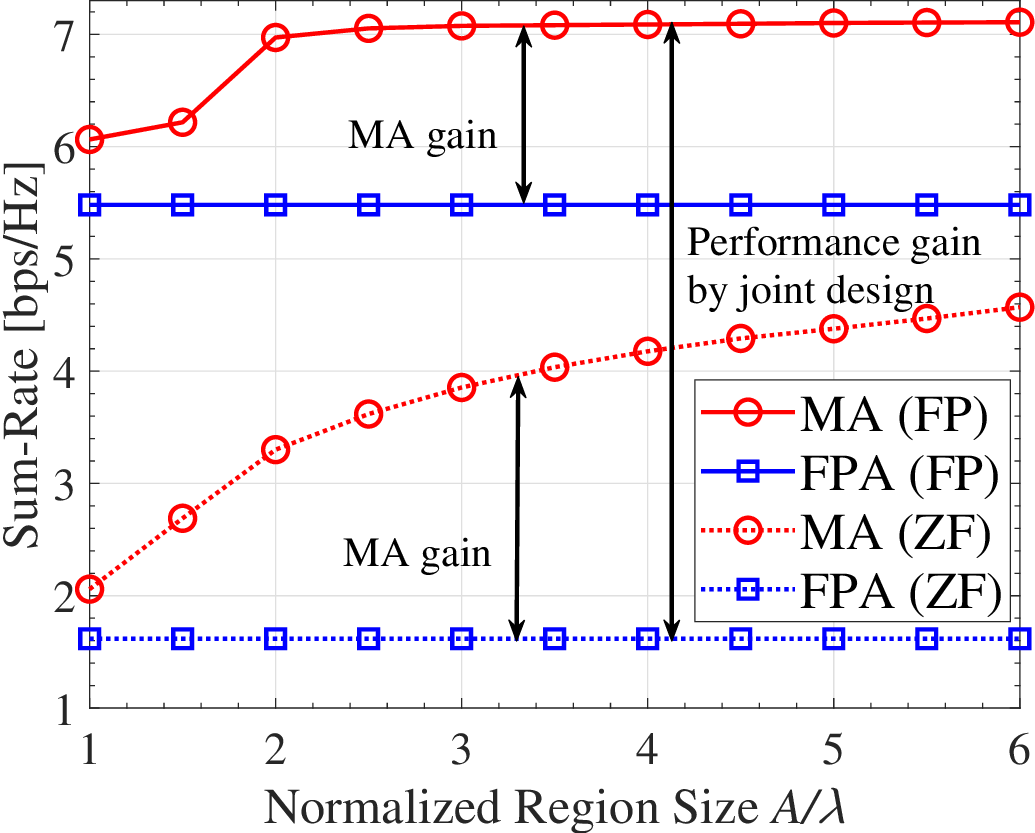}
\caption{Sum-rate vs. the normalized region size. $p=10$ dBm.}
\label{fig2b}
\vspace{-10pt}
\end{figure}
In {\figurename} {\ref{fig2b}}, we show the achievable sum-rate versus the normalized region size $A/\lambda$. It is observed that the proposed schemes with MAs outperform FPA systems in terms of achievable rate, and the performance gain increases with the region size. It is also observed that our proposed FP-based design, i.e., Algorithm \ref{Algorithm2}, achieves the best performance among all schemes for any region size. Furthermore, it is worth highlighting that for MAs, the FP-based scheme converges when the normalized region size is larger than 3. This suggests that the optimal sum-rate performance for MA-enabled communication systems can be achieved within a finite transmit region. These findings underscore the efficacy of our proposed algorithms in reshaping the multiuser channel to create a more favorable environment for the maximization of the sum-rate.
\section{Conclusion}
In this letter, we introduced a multiuser transmission system empowered by MAs with the aim of enhancing the overall sum-rate performance through antenna position optimization. Our investigation focused on the joint optimization of transmit beamforming and transmit MA positions. We presented a pair of efficient algorithms, leveraging the principles of alternating optimization, gradient descent, and backtracking line search methods. Numerical results revealed that the proposed MA-based architecture provides more DoFs for improving the sum-rate and outperforms conventional FPA-based ones.

\end{document}